\documentclass{llncs}

\usepackage{epsf}

\newcommand{\eq}{\begin{equation}}
\newcommand{\en}{\end{equation}}
\newcommand{\eqa}{\begin{eqnarray}}
\newcommand{\ena}{\end{eqnarray}}
\begin{document}

\pagestyle{empty}

\mainmatter

\title{Synchronisation, binding and the role of correlated firing in fast information transmission}

\titlerunning{Correlations and Information}

\author{Simon R. Schultz$^{1}$
\and Huw D. R. Golledge$^{2}$ \and Stefano Panzeri$^{2}$}

\authorrunning{Simon R. Schultz et al.}

\institute{
Howard Hughes Medical Institute \& Center for Neural Science, New York University, New York, NY 10003, USA
\and
Department of
Psychology, Ridley Building, \\ University of Newcastle upon Tyne, Newcastle NE1 7RU, UK \\
\email{stefano.panzeri@ncl.ac.uk}\\
\texttt{http://www.staff.ncl.ac.uk/stefano.panzeri/}
}

\maketitle

\begin{abstract}
Does synchronization between action potentials from different neurons
in the visual system play a substantial role in solving the binding
problem? The binding problem can be studied quantitatively in the
broader framework of the information contained in neural spike trains
about some external correlate, which in this case is object
configurations in the visual field. We approach this problem by using
a mathematical formalism that quantifies the impact of correlated
firing in short time scales. Using a power series expansion, the
mutual information an ensemble of neurons conveys about external
stimuli is broken down into firing rate and correlation
components. This leads to a new quantification procedure directly
applicable to simultaneous multiple neuron recordings. It
theoretically constrains the neural code, showing that correlations
contribute less significantly than firing rates to rapid information
processing. By using this approach to study the limits upon the amount
of information that an ideal observer is able to extract from a
synchrony code, it may be possible to determine whether the available
amount of information is sufficient to support computational processes
such as feature binding.
\end{abstract}


\section{Introduction}

Does synchronization (or more generally temporal correlations) between
action potentials from different cells in the central visual system play
a substantial role in solving crucial computational problems, such as
binding of visual features or figure/ground segmentation? One theory
suggests that synchrony between members of neuronal assemblies is the
mechanism used by the cerebral cortex for associating the features of
a coherent single object \cite{VdM95}.

Although several groups have reported compelling experimental evidence
from the visual system in support of this theory (for a review see
\cite{Sin+97}), the role played by synchronous firing in visual
feature binding is still highly controversial
\cite{Shadlen99,Ghose99,You+92,Golledge96}, and far from being understood. In
our view, it is possible that one or more methodological factors
contribute to the continuing uncertainty about this issue. In fact,
almost all the reported neurophysiological evidence in favor or
against the temporal binding hypothesis relies upon the assessment of
the significance of peaks in cross-correlograms (CCG, \cite{Aer+89})
and of their modulation with respect to stimulus configuration. While
investigating stimulus modulation of peaks (or of other features) of
CCGs can clearly bear some evidence on the role of synchrony in
binding, it does not address the crucial issue of {\em how much}
synchrony tells the brain about the configuration of objects in the
visual field. This question is particularly important as it is well
known that firing rates of individual cells are commonly related to
features of the sensory world \cite{Adr26}, and even to perceptual
judgements (see e.g. \cite{Bri+92,Shadlen96pnas}). Firing rate modulations
can potentially contribute to association of features through the use
of population codes, or also in other ways\cite{Shadlen99}. Therefore the specific contribution of synchrony (or in
general of correlations between firing of cells) as a coding mechanism
for binding should be assessed against the contribution of independent
firing rate modulation to the encoding of object configurations in the
visual field.

To address these issues, a pure analysis of CCG characteristics is
insufficient. In addition to CCG quantification, information theory
can be used to address the specific contribution of synchronized or
correlated firing to visual feature binding, and to compare the
contribution of synchrony against that of firing rates. In fact,
Information theory \cite{Sha48} allows one to take the point of view
of an ideal observer trying to reconstruct the stimulus configuration
just based on the observation of the activity of neuronal population,
and to determine how much the presence of correlated firing helps in
identifying the stimulus.

In this paper we present and develop a rigorous information theoretic
framework to investigate the role of temporal correlations between
spikes. We first discuss how information theory could overcome the
limitations of the pure CCG analysis. We then present a mathematical
formalism that allows us to divide the information into components
which represent the information encoding mechanisms used by
neuronal populations -- i.e. it determines how many bits of
information were present in the firing rates, how many in coincident
firing by pairs of neurons, etc., with all of these adding up to the
overall available information. The mathematical approach developed
here is valid for timescales which are shorter than or of the order of a
typical interval between spike emissions, and it makes use of a Taylor
series approximation to the information, keeping terms up to the
second order in the time window length. This approximation is not
merely mathematically convenient; short timescales are likely to be of
direct relevance to information processing by the brain, as there is
substantial evidence that much sensory information is transmitted by
neuronal activity in very short periods of time
\cite{Tov+93,Tho+96,Rol+99mask}. Therefore the mathematical analysis
is relevant to the study of the computations underlying perceptual
processes. In particular, it enables the quantitative study of the
rapidly appearing correlational assemblies that have been suggested to
underlie feature binding and figure/ground segmentation.

\section{Problems with conventional Cross-Correlogram analysis}

The CCG represents a histogram of the probability of a spike from one
neuron at a given time relative to a reference spike of a second neuron \cite{Aer+89}. Whilst cross-correlation is capable of identifying synchrony between neurons, several aspects of the analysis of CCGs present problems or are incomplete. First, CCG analysis itself does not provide a criterion to choose which periods
of a response epoch should be analysed. Since, in many cases, moving
stimuli are employed, the response varies with time and it may be that
correlations are present or are stimulus modulated for only a short
part of the response \cite{Gra+92}. This short period is not
necessarily related simply to the response peak, although some studies
have analysed only the period in which the peak of the response is
made to a moving stimulus \cite{Kreiter96}.
Second, the width of the
time window over which correlations should be assessed is arbitrary. CCG analysis does not entirely address over which time scales correlations contribute most
information about object configuration.
Using long windows (e.g. much larger than the width of CCG peaks) may
``wash out'' transient correlations. Narrow windows centered upon the
PSTH peak may ignore the part of the responses that contains most of
the information about the stimuli (e.g. in firing rate
modulations). Third, if the window length used to assess correlations
is varied between stimulus conditions (e.g \cite{Kreiter96}) then an
undesirable extra source of variation is introduced when the stimulus
conditions are compared.
Information theory can mitigate some of these problems by providing a
criterion for the selection of time windows, by identifying the
windows in which most information is actually transmitted.

Many
previous studies also differ in the methods used to quantify the
temporal structure in CCGs. Some studies rely on the fitting of a
damped sine wave to the CCG (e.g. \cite{You+92,Kon+95pnas}). Other methods quantify solely the likelihood that the peak in the CCG did not arise by chance
\cite{Oliveira97}. Analysis of the significance of a peak, or of structure
in the CCG must be made in relation to the flanks of the CCG. What
length of flank is chosen will affect the significance of
peaks. However, downstream neurons are unlikely to be able to compare
the likelihoods of spikes occurring at lags of tens of milliseconds
against the likelihood of simultaneous spikes.

The parameters of a CCG do not themselves quantify the informational
contribution of synchronous firing. Conventional CCG analysis
techniques attempt to assess correlation in a manner independent of
the firing rate in order to disambiguate synchronous modulations from
firing rate variations. It is unlikely, though, that any downstream
detector of the synchronous discharge of neurons would be capable of
assessing the significance of correlation independent of the firing
rate. It is more likely that it would make use of the actual number of
coincident spikes available in its integration time window. Therefore
cross-correlation peaks and firing rate modulation are probably
intrinsically linked in transmitting information to downstream
neurons, and an analysis of the functional role of synchrony should be
able to take this into account. Most studies that appear to show
stimulus-dependent synchrony have employed relatively strongly
stimulated cells. An important prediction of the temporal correlation
hypothesis is that synchrony should encompass the responses of
sub-optimally stimulated neurons \cite{Konig95}. A thorough test of
this hypothesis requires the study of cells that fire very few spikes.
The number of spikes included in the calculation of a CCG of course
affects the precision with which correlations can be detected
\cite{Hata91}. Variations in the number of evoked spikes, rather than
a true change in correlation between the neurons, could thus affect
comparisons between optimal and sub-optimal stimuli. While analysing
cells firing at low rates may be a challenge for CCG analysis, it is
tractable for analyses developed from information theory, as we shall see in Section 4.

\section{Information Theory and Neuronal Responses}

We believe that the methodological ambiguities that attend
studies purely based on quantification of spike train correlograms can be greatly reduced by employing in addition methods based upon information theory \cite{Sha48}, as we describe in this Section.

Information theory \cite{Sha48} measures the statistical significance
of how neuronal responses co-vary with the different stimuli presented
at the sensory periphery. Therefore it determines how much
information neuronal responses carry about the particular set of
stimuli presented during an experiment. Unlike other simpler measures,
like those of signal detection theory, which take into account only
the mean response and its standard deviation, information theory
allows one to consider the role of the entire probability
distributions. A measure of information thus requires sampling
experimentally the probabilities $P(r|s)$ of a neuronal population
response $r$ to all stimuli s in the set, as well as designing the
experimental frequencies of presentation $P(s)$ of each stimulus. The
information measure is performed by computing the distance between the
joint stimulus-response probabilities $P(r,s)= P(r|s) P(s)$ and the
product of the two probabilities $P(r)P(s)$, ($P(r)$ being the
unconditional response probability) as follows:

\begin{equation}
I(S;R) = \sum_s \sum_r P(s,r) \log_2 {P(s,r) \over P(s) P(r)}
\label{infodef}
\end{equation}

If there is a statistical relationship between stimuli and responses
(i.e. if $P(r,s)$ is dissimilar from $P(r)P(s)$) , our knowledge about
what stimulus was presented increases after the observation of one
neuronal spike train. Eq. (\ref{infodef}) quantifies this fact. The
stronger the statistical relationship between stimuli and responses,
the higher is the information value. Eq. (\ref{infodef}) thus
quantifies how well an ideal observer could discriminate between
different stimulus conditions, based on a single response trial. There
are several advantages in using information theory to quantify how
reliably the activity of a set of neurons encodes the events in the
sensory periphery \cite{Rie+96,Borst99}. First, information theory puts
the performance of neuronal responses on a scale defined at the ratio
level of measurement. For example, an increase of 30\% of on the peak
height of a cross-correlogram does not tell us how this relates to
synchrony-based stimulus discrimination, but values of information
carried by synchronous firing have a precise meaning. Information
theory measures the reduction of uncertainty about the stimulus
following the observation of a neuronal response on a logarithmic
scale. One bit of information corresponds to a reduction by a factor
of two in the stimulus uncertainty. A second advantage of information
theory in this context is that it does not require any specific
assumption about the distance between responses or the stationarity of
the processes, and it can therefore lead to objective assessments of
some of the hypotheses.

\begin{figure}
\begin{center}
\leavevmode
\epsfysize=10truecm
\epsfxsize=10truecm
\epsffile{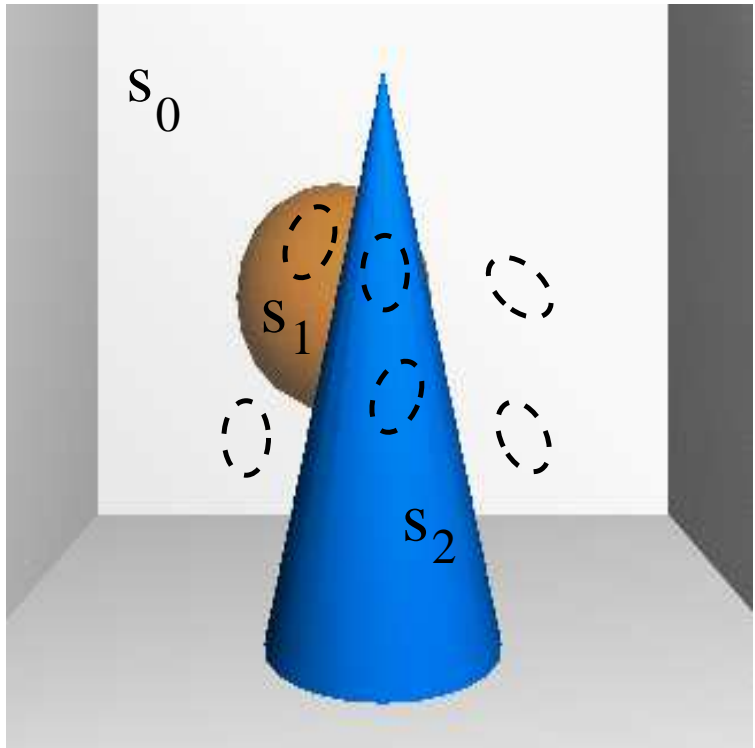}
\end{center}
\caption{An illustrative segregation problem in which there are two objects in front of a background. The background is labeled by $s_0$, and the objects by $s_1$ and $s_2$ respectively. The dashed ellipses represent the receptive fields of visual cortical cells which we are recording the responses of simultaneously. This situation can be examined in the framework of information theory by considering the `stimulus' to be a multidimensional variable indicating which object ($s_0$, $s_1$ or $s_2$) is associated with each receptive field. The problem is thus to determine which response characteristics are most informative about the visual configuration.}
\label{fig:ill}
\end{figure}

In the above discussion we have mentioned that we are calculating
information `about a stimulus'. In fact, more generally it can be
information `about' any quantifiable external correlate, but we shall
continue to use the word stimulus in an extended sense. If we were
studying information about the orientation of a grating, we would
define our stimulus to be which of a number of different orientations
the grating appeared at on any given experimental trial. If we wish to
study problems such as binding or figure-ground segregation within
this framework, we have to specify our stimulus
description accordingly. An illustration of this is shown in the scene of
Fig.~\ref{fig:ill}, which contains two objects in front of a
background. Also shown are a number of receptive fields, which are
taken to be those of cells from which we are simultaneously recording
the activity (`spike trains'). We can define our stimulus, or external
correlate, as a multidimensional variable representing the object to
which each receptive field is associated. The dimensionality of our
stimulus is in this case the number of cells from which we are
recording at once. By quantifying the information contained in the
spike trains about this variable, and breaking it down into individual
components reflecting firing rates and correlations (or
synchronisation), we can determine the aspects of the spike train
which best encode the figure-ground (or object-object-ground)
segregation. Furthermore, by examining how this relationship scales
with the stimulus dimensionality (number of receptive fields recorded
from), it may be possible to determine whether enough information is
present in correlations to support binding in perceptually realistic
environments.

It is worth noticing that information values are always relative to
the stimulus set used, and that testing a neuron with different
stimuli may lead to rather different information values. This has some
interesting implications. On the one hand, it allows us to
characterise neuronal selectivity by searching for a stimulus set that
maximises the neuronal information transfer, a more rational
characterisation strategy than searching for stimuli eliciting
sustained responses. On the other hand, the intrinsic dependency of
mutual information values on the nature of stimulus set allows us to
test whether different encoding strategies are used by visual cortical
neurons when dealing with external correlates of a different
nature. The last property is of interest because one of the
predictions of the binding-by-synchrony hypothesis is that synchrony
is particularly important when stimulus configurations requiring some
kind of associations are included, and less important in other
situations. Information theory thus provides a natural framework to
test this theory.

\section{Series expansion of the mutual information}

Although information theory is, as explained, a natural framework to
address the role of correlated firing on e.g. binding of visual
features, some work is needed to separate out of the total information
contained in a population response $r$ into components, each
reflecting the specific contribution of an encoding mechanism. We
perform this separation in the limit in which the relevant window for
information encoding is short enough that the population typically
emits few action potentials in response to a stimulus. As discussed in
the introduction and in \cite{Pan+99cor}, there is evidence that this
is a relevant limit for studying the computations underlying
perception, as cortical areas in several cases perform their
computations within a time frame which is shorter than one typical
interspike interval \cite{Tho+96,Rol+99mask}.

We examine a period of time of duration $t$, in which a stimulus $s$
is present. The neuronal population response $r$ during this time is
taken to be described by the number of spikes fired by each
cell\footnote{The additional temporal information contained in the
spike times is studied in \cite{Panzeri00temp}} in the window $t$. We study the information
carried by the responses of a neuronal population about which stimulus
is presented. It can be approximated by a power series
\begin{equation}
I(t) = t\; I_t +
\frac{1}{2} t^2\; I_{tt} + O(t^3).
\label{taylor}
\end{equation}
 The problem is thus to compute the
first two time derivatives of the information, $I_t$ and $I_{tt}$, which
are all that survive at short timescales.

Two kinds of correlations influence the information. These are the
``signal'' (mean response to each stimulus) correlation and the ``noise''
(variability from the mean across different trials with the same stimulus)
correlation between cells. In the short timescale limit the noise
correlation can be quantified as
\eq
\gamma_{ij}(s) = \frac{\overline{n_{i}(s)
n_{j}(s)}}{(\overline{n}_{i}(s)\overline{n}_{j}(s))} -1 , (i \neq j)
\qquad
\gamma_{ii}(s) = \frac{(\overline{n_{i}(s)^2} - \overline{n_{i}(s)})}{
\overline{n_{i}(s)}^2} - 1,
\en
where $n_i(s)$ is the number of spikes emitted by cell $i$ in a given
trial in response to stimulus $s$, and the bar denotes an average
across experimental trials to the same stimulus. The signal
correlation can be measured as
\begin{equation}
\nu_{ij} = { < \overline{r}_i(s) \overline{r}_j(s)>_s \over
<\overline {r}_i(s)>_{s} <\overline {r}_j(s)>_{s} } -1,
\label{signalcorr}
\end{equation}
where $\overline{r}_i(s)$ is the mean firing rate of cell $i$ to stimulus $s$, and $<\cdots>_s$ denotes an average across stimuli.
These are scaled correlation densities ranging from -1 to
infinity, which remain finite as $t \to 0$.
Positive values of the correlation coefficients indicate positive correlation, and negative values indicate anti-correlation.

Under the assumption that the probabilities of neuronal firing conditional upon the
firing of other neurons are non-divergent, the $t$
expansion of response probabilities becomes an expansion in
the total number of spikes emitted by the population in response to a
stimulus. The probabilities of up to two spikes being emitted are
calculated and inserted into the expression for information. This yields
for the information derivatives
\eqa I_t &=& \sum_{i=1}^C
\left< \overline{r}_{i}(s) \log_2 {\overline{r}_{i}(s)\over
\left<\overline{r}_{i}(s')\right>_{s'}}~\right>_{s}\\
\label{eq:1stderiv}
I_{tt} & = & {1\over \ln 2} \sum_{i=1}^C  \sum_{j=1}^C
\left<\overline {r}_{i}(s)\right>_{s} \left<\overline
{r}_{j}(s)\right>_{s}  \biggl[
\nu_{ij} + (1 + \nu_{ij})\ln ({1\over 1+\nu_{ij}} ) \biggr]
\nonumber \\
& + &  \sum_{i=1}^C  \sum_{j=1}^C \biggl[ \left< \overline{r}_{i}(s)
\overline{r}_{j}(s)
\gamma_{ij}(s) \right>_s
\biggr] \log_2 ({1\over 1+\nu_{ij}}) \nonumber \\
& + & \sum_{i=1}^C  \sum_{j=1}^C \left< \overline{r}_{i}(s)
\overline{r}_{j}(s) (1 + \gamma_{ij}(s))
\log_2 \biggl[{(1+ \gamma_{ij}(s)) \left<\overline{r}_{i}(s')
\overline{r}_{j}(s')\right>_{s'}\over
\left<\overline{r}_{i}(s')
\overline{r}_{j}(s')(1+ \gamma_{ij}(s'))\right>_{s'} } \biggr] \right>_s .
\label{eq:2ndderiv}
\ena
The first of these terms is all that survives if there is no noise
correlation at all.  Thus the {\em rate component} of the information is
given by the sum of $I_t$ (which is always greater than or equal to zero)
and of the first term of $I_{tt}$ (which is instead always less than or
equal to zero). The second term is non-zero if there is some correlation
in the variance to a given stimulus, even if it is independent of which
stimulus is present; this term thus represents the contribution of {\em
stimulus-independent noise correlation} to the information.  The third
component of $I_{tt}$ is non-negative, and it represents the contribution of {\em
stimulus-modulated noise correlation}, as it becomes non-zero only for
stimulus-dependent correlations. We refer to these last two terms of
$I_{tt}$ together as the correlational components of the information.

In any practical measurement of these formulae, estimates of finite
sampling bias must be subtracted from the individual
components. Analytical expressions for the bias of each component are
derived in the online appendix of \cite{Pan+99cor}.

\section{Correlations and fast information processing}

The results reported above for the information derivatives show that
the instantaneous rate of information transmission (related to the
leading order contribution to the information) depends only upon the
firing rates. Correlations contribute to information transmission, but
they play only a second order role. This has interesting implications
for the neural code, that we develop further here.

It was argued \cite{Sha+98} that, since the response of cortical
neurons is so variable, rapid information transmission must imply
redundancy (i.e. transmitting several copies of the same message).  In
other words, it should necessary to average away the large observed
variability of individual interspike intervals s by replicating the
signal through many similar neurons in order to ensure reliability in
short times.  Our result, eq. (\ref{eq:1stderiv}), shows that to have a
high information rate, it is enough that each cell conveys some
information about the stimuli (because the rate of the information
transmitted by the population is the sum of all single cell
contributions); therefore we conclude that it is not necessary to
transmit many copies of the same signal to ensure rapidity.

Also, since firing rates convey the main component of information,
correlations are likely to play a minor role in timescales of the
order of 20-50 ms, in which much information is transmitted in the
cortex. As an example, Fig. 2 shows the information conveyed by a
population of simulated Integrate and Fire neurons, which share a
large proportion (30\%) of their afferents (see \cite{Pan+99cor} for
details of the neuronal model used). It can be seen that, despite the
strong correlations between the cells in the ensemble correlations
play only a minor role with respect to firing rates.

\begin{figure}
\begin{center}
\leavevmode
\epsfysize=6truecm
\epsfxsize=6.5truecm
\parbox[t]{6.8cm}{\bfseries\Large a\epsffile{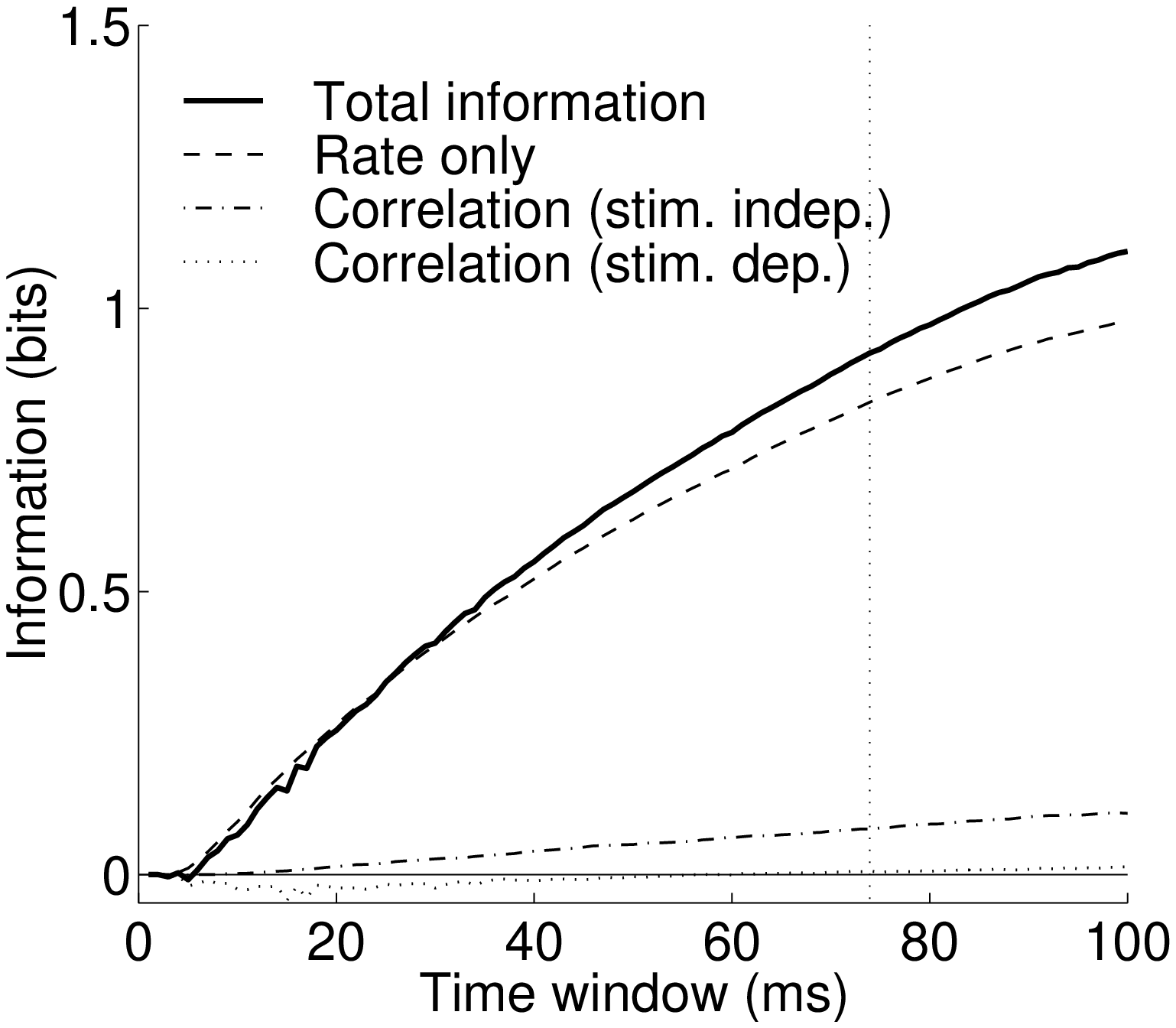}}
\epsfysize=6truecm
\epsfxsize=6.5truecm
\parbox[t]{6.8cm}{\bfseries\Large b\epsffile{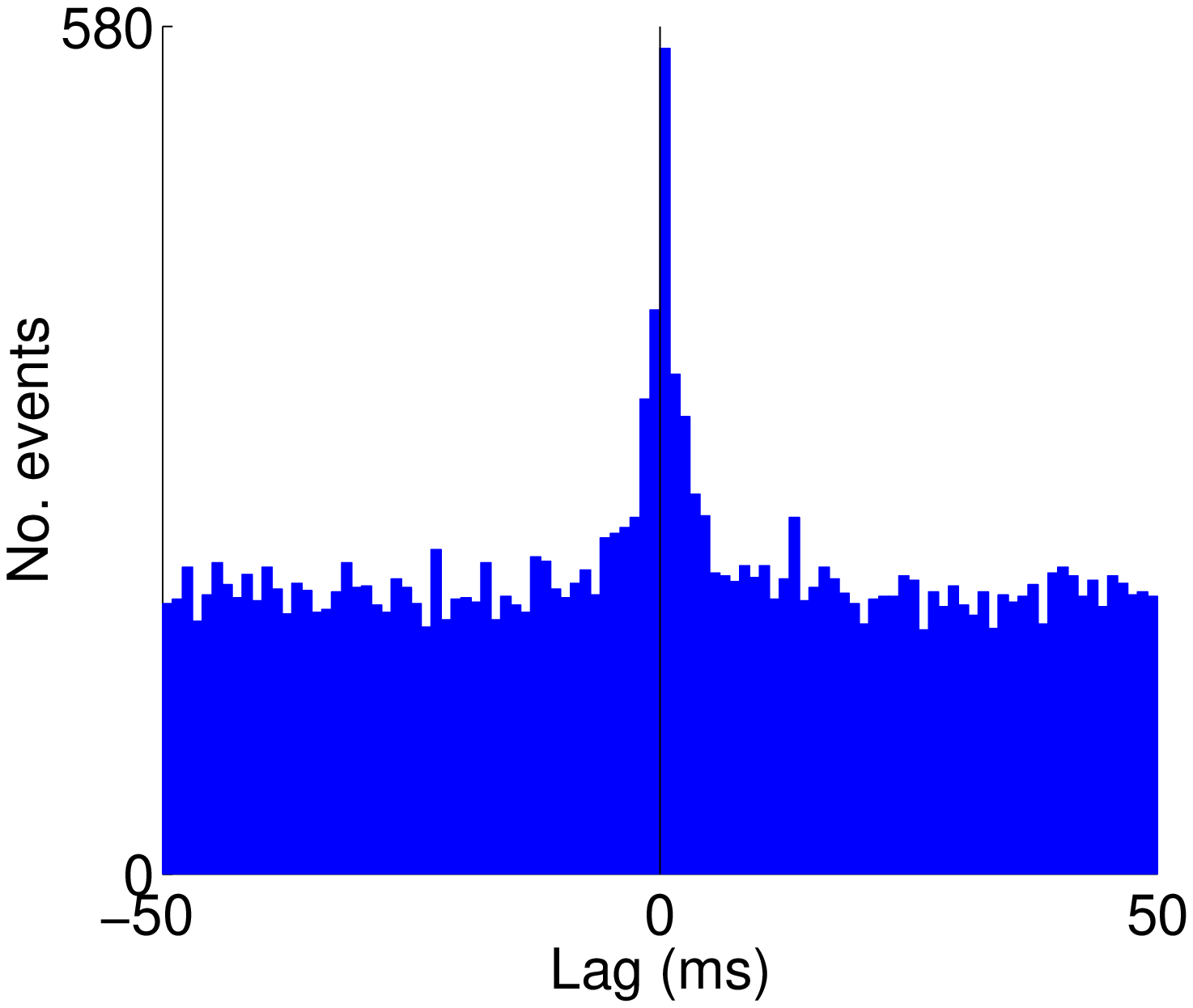}}
\end{center}
\caption{ (a) The short-timescale information components for a set of 5 simulated neurons sharing 30\% of their inputs. (b) A typical example of the cross-correlogram between two of the simulated neurons. Despite of the strong correlation between neurons, the impact of the cross correlations on information transmission is minimal.}
\end{figure}

To model a situation where stimulus dependent correlations conveyed information, we generated simulated data using the Integrate-and-Fire model for another quintuplet of cells which had a stimulus dependent fraction of common input.
This might correspond to a situation where transiently participate in different neuronal assemblies, depending on stimulus condition. This is therefore a case that might be found if the binding-by-synchrony theory is correct.
There were ten stimuli in the sample. The spike emission rate was constant (20 Hz) across stimuli.  One of the stimuli resulted in independent input to each of the model cell, whereas each of the other nine stimuli resulted in an increase (to 90\%) in the amount of shared input between one pair of cells. The pair was chosen at random from the ensemble such that each stimulus resulted in a different pair being correlated The change in responses of one such pair of cells to changes in the amount of common input is shown in Fig.~3a.
The upper panel of Fig.~3a shows the fraction of shared connections as a function of time; the central and lower panel of Fig. 3a show the resulting membrane potentials and spike trains from the pair of neurons. This cross-correlation is also evident in the cross-correlograms
shown in Fig.~3b. The results for the information are given in Fig.~3c: all terms but the third of $I_{tt}$ are essentially zero, and information transmission is in this case almost entirely due to stimulus-dependent
correlations. This shows that the short time window series expansion pursued here is able to pick up the right encoding mechanisms used by the set of cells.
Therefore it is a reliable method for quantifying the information carried by the correlation of firing of small populations of cells recorded from the central nervous system {\em in vivo}. Another point that is worth observing is that Fig.~3c also shows that the total amount of information that could be conveyed,
even with this much shared input, was modest in comparison to that
conveyed by rates dependent on the stimuli, at the same mean firing
rate. This again illustrates that correlations typically convey less information than what can be normally achieved by rate modulations only. Therefore they are likely to be a secondary coding mechanism when information is processed in time scales of the order of one interspike interval.

\begin{figure}
\begin{center}
\leavevmode
\epsfysize=4truecm
\epsfxsize=4.5truecm
\parbox[t]{4.8cm}{\bfseries\Large a\epsffile{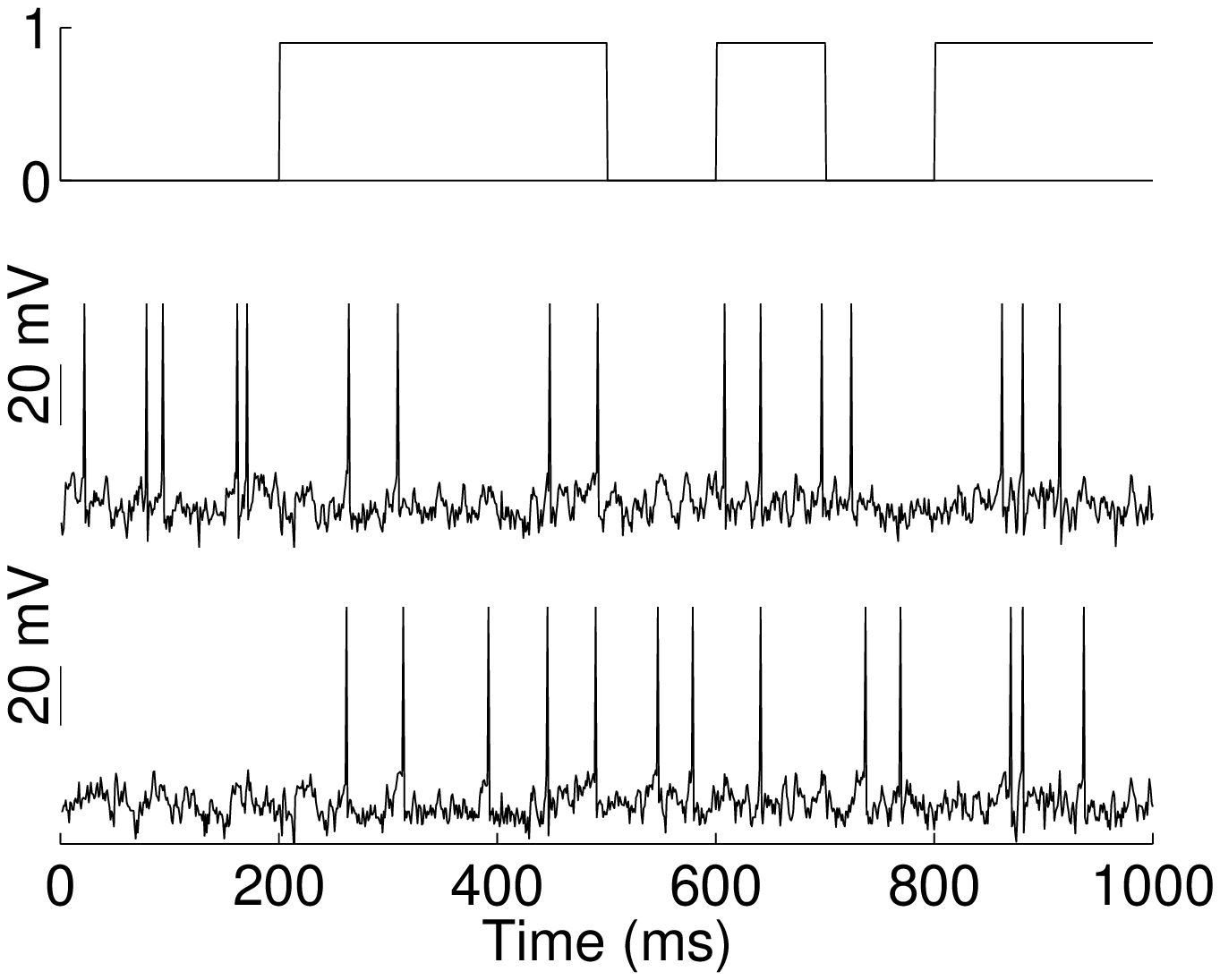}}
\epsfysize=4truecm
\epsfxsize=4.5truecm
\parbox[t]{4.8cm}{\bfseries\Large b\epsffile{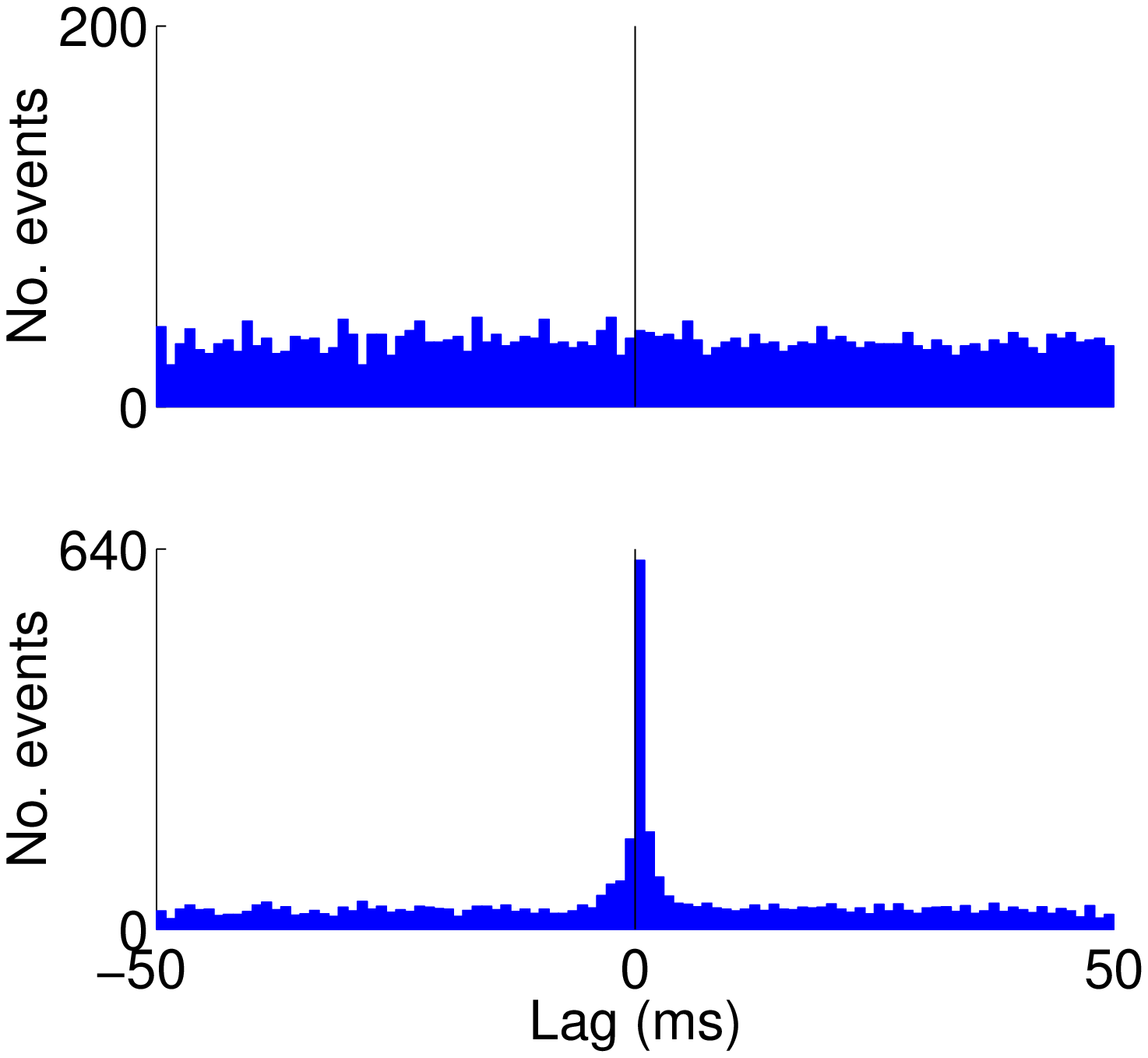}}
\end{center}
\begin{center}
\leavevmode
\epsfysize=8truecm
\epsfxsize=8truecm
\parbox[t]{8.5cm}{\bfseries\Large c\epsffile{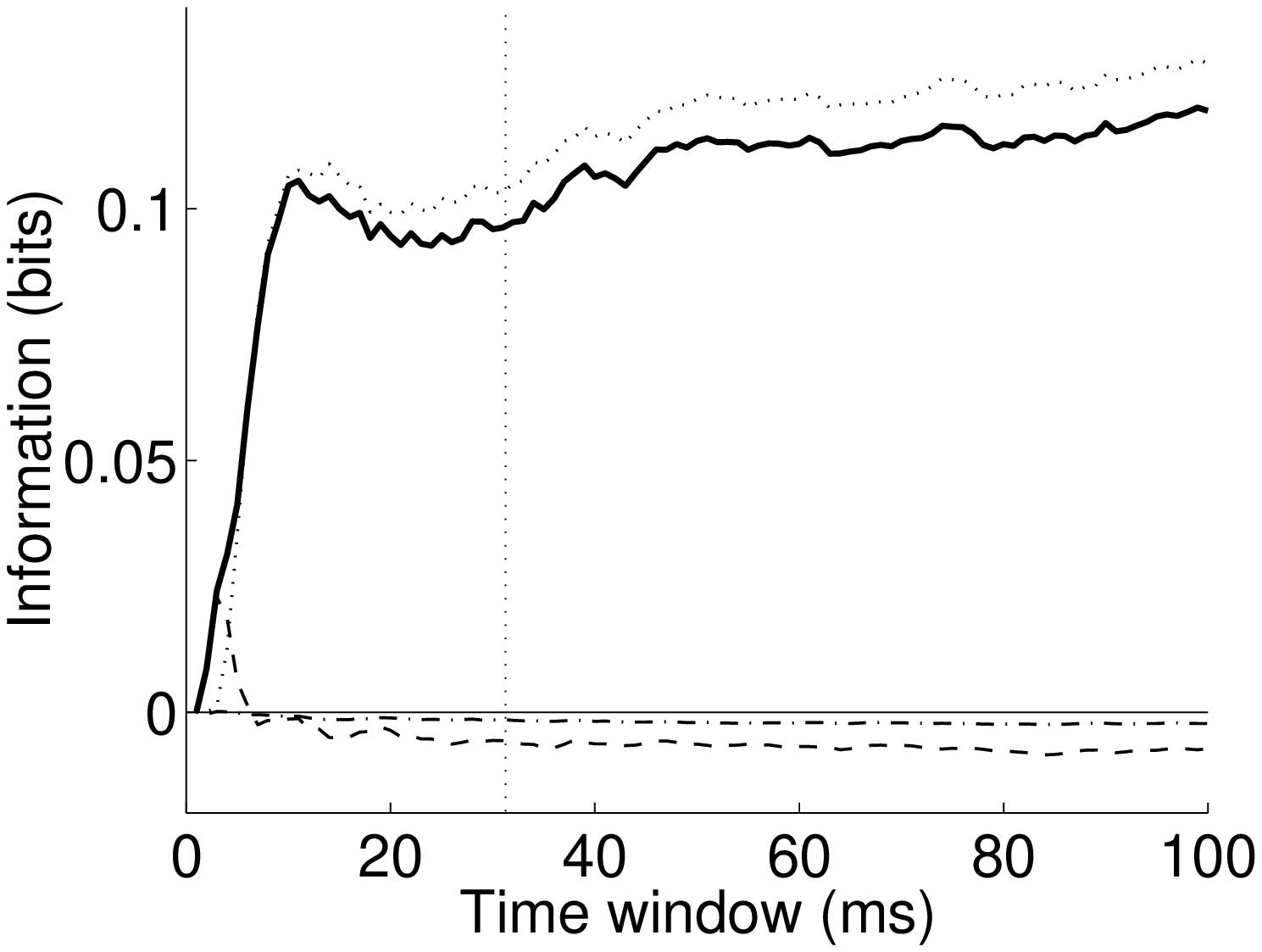}}
\end{center}
\caption{A situation in which the stimulus dependent correlational component dominates: with a fixed mean firing rate, two of the five simulated cells
(chosen randomly for that stimulus) increase their correlation by
increasing the number of shared connections while the other two remained
randomly correlated. The effect of this on cell spiking activity is shown
in (a): upper panel shows the fraction of shared connections, while central and lower panels of (a) show the membrane potential and spike emission of the
simulated cells. (b) shows the cross-correlograms corresponding to the low
and high correlation states. The result of this is seen in (c):
information due to correlations, although modest in magnitude, in this
demonstration dominates the total information.}
\end{figure}

\section{Optimality of correlational encoding}

In the preceding sections we have shown that the correlational
component is only second order in the short time limit, essentially
because the probability of emission of pairs of spikes, and the
reliability of this process, are much smaller than the corresponding
quantities for single spikes. For this reason, one can expect a
correlational code to carry appreciable information only when it is
{\em efficient}, i.e. when each correlated spike pair carries as much
information as possible.

In this section we investigate the statistical conditions that have to be
satisfied by a correlational code in order to
be efficient in the short time limit. If the population code is purely correlational
(i.e. the firing rates are not modulated at all by the stimuli), then
it is possible to show that the mean information per coincidence $\Psi$ carried
by a pair of cells (obtained dividing the total information by the
mean number of observed coincidences) is bounded only by the
sparseness of the distribution of coincident firing across stimuli
$\alpha$:

\eq 0 \leq
\Psi \leq \log_2(1/\alpha)
\label{ineqPsi}
\en

The maximal (most efficient) value of information per coincidence
$\Psi_{\rm max} = \log_2(1/\alpha)$ is reached by a binary distribution of correlations
across stimuli, with a fraction of stimuli $\alpha$ eliciting
positively correlated firing, and the other $1 - \alpha$ stimuli
eliciting fully anti-correlated firing (i.e. coincident firing is never
observed when presenting one of the latter stimuli). Nearly uniform,
or strongly unimodal distributions of correlations across stimuli
would give poor information, $\Psi \sim 0$.

By analyzing Eq. (\ref{ineqPsi}), it is easy to realize that there are two statistical requirements that are necessary to achieve high values of information per coincidence. The first one is that the correlational code should be {\ em sparse} (i.e. the fraction of stimuli leading to a ``high correlation state'' should be low). The sparser the code, the more information per coincidence can be transmitted.  The second important factor
for fast and efficient transmission of correlational information, is
that the low correlational state must be strongly {\em anti-correlated} in order to achieve an information per coincidence close to its maximum
$\log_2(1/\alpha)$. In fact correlations in short times have
fluctuations that may be big compared with their mean value, and
therefore for any observer it is difficult to understand in less than
one ISI if an observed coincidence is due to chance or neuronal
interaction. This is why low correlational states with no coincidences
are so helpful in transmitting information. We note here that states
of nearly complete anticorrelation have never been observed in the
brain. Therefore the ``low state correlational state'' of a realistic
correlational assembly should be the ``random correlation state'' (i.e. the state in which the number of coincident spikes is on average that expected by chance).

We have quantified the reduction in the information per coincidence, compared to its maximum $\Psi_{\rm max}$, that arises as a consequence of the presence of the random correlation state.
Fig.~4 plots the ratio between the information per coincidence carried when the ``low correlation state'' is random and the optimal amount of information per coincidence $\log_2(1/\alpha)$ obtained when the low correlation state is totally anticorrelated. The plot is shown as a function of the fraction $\alpha$ of stimuli eliciting a ``highly correlated state''. Fig.~4 shows clearly that, if the ``low correlation state'' of the assembly
elicits uncorrelated firing, then the information per coincidence is
far from its maximum, unless the correlation in the ``high state'' is
extremely high and the correlational code is not sparse at all. However, in
which case the information per coincidence is very low anyway (see eq. (\ref{ineqPsi})).

Therefore correlational assemblies in
the brain, if any, are likely to be inefficient in the short time
limit. This consideration further limits the possible role played by
correlations in fast information encoding.

\begin{figure}
\begin{center}
\leavevmode
\epsfysize=6truecm
\epsfxsize=6.5truecm
\parbox[t]{6.8cm}{\epsffile{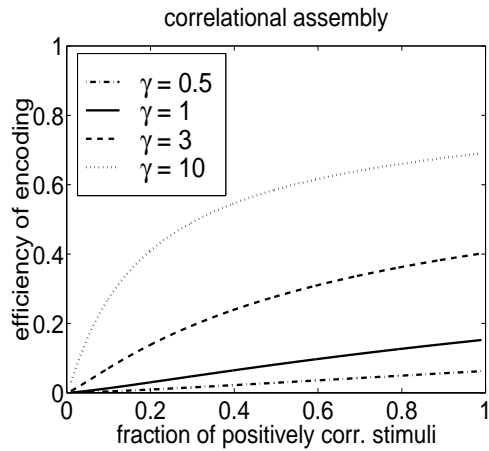}}
\end{center}
\caption{The ratio between the information per coincidence carried by
a binary correlational encoder with a fraction of stimuli eliciting
positive correlation $\gamma$ and the other stimuli
eliciting no correlation, and the optimal information per coincidence
carried in the same situation, but with full anticorrelation in the
``low correlation state''. This ratio is plotted, for different values
of the strength $\gamma$ of the ``high correlation state'', as a function of the fraction of stimuli eliciting positive correlation.}
\end{figure}

\vspace{-1cm}
\section{Discussion}

If cells participate in context-dependent correlational assemblies
\cite{Sin+97}, then a significant amount of information should be
found in the third component of $I_{tt}$ when analysing data obtained
from the appropriate experiments. The series expansion approach thus
enables the testing of hypotheses about the role of correlations in
solving the binding problem, as opposed to other solutions, and about
information coding in general. Data analyses based on the
time-expansion approach have the potential to elucidate the role of
correlations in the encoding of information by cortical neurons.

It is worth noticing that the formalism presented here evaluates the
the information contained in the neuronal responses themselves, it
does not make assumptions about the system that is going to read the code. For
this reason, the information computed ``directly'' from neuronal
responses is an upper bound to what any type of decoder can extract
from the responses themselves. Therefore it is termed the information
that an ``ideal observer'' can extract \cite{Borst99}. Of course, the
relative contribution of rate and synchrony modulations to information
transmission will depend on the specific read-out mechanism used by a
downstream neural system that listens to the neuronal code. However,
if the information that an ideal observer is able to extract from the
synchrony code is small, as the mathematical analysis indicates for
the fast information processing limit, one can be sure that any
decoding device cannot extract more information in the synchrony than
that small amount evaluated from the responses.

Whether this small amount is sufficient to support computational
processes such as figure-ground segregation remains to be seen, and
depends upon how it scales empirically with the number of receptive
fields examined. Ultimately, as suggested by \cite{Shadlen99}, the only
way we will achieve any degree of confidence in a proposed solution to
the binding problem will be to study recordings made from a monkey
trained to make behavioural responses according to whether individual
features are bound to particular objects. The information theoretic
approach described here would be a natural way to analyse such data.

In conclusion, the methodology presented here can provide interesting and
reliable bounds on the role of synchrony on cortical information
encoding, and we believe that its application to neurophysiological
experiments will advance our understanding of the functional interpretation
of synchronous activity in the cerebral cortex.

\section{Acknowledgements}
We would particularly like to thank Alessandro Treves and Malcolm Young for useful discussions relating to this work.

\bibliographystyle{unsrt}
\bibliography{srs}
\end{document}